\documentstyle[multicol,prl,aps,epsf] {revtex}
\begin{document}
\bibliographystyle{prlsty}
\draft
\title{Vanishing of the negative-sign problem of quantum Monte Carlo 
       simulations in one-dimensional frustrated spin systems}
\author{Tota Nakamura}
\address{Department of Applied Physics, Tohoku University,
         Sendai, Miyagi 980-77, Japan}
\date{\today}
\maketitle
\begin{abstract}
The negative-sign problem in 
one-dimensional frustrated quantum spin systems is solved.
We can remove negative signs of the local Boltzmann weights
by using a dimer basis that has the spin-reversal symmetry.
Validity of this new basis is checked in a general frustrated 
double-spin-chain system, namely the $J_0$-$J_1$-$J_2$-$J_3$ model.
The negative sign vanishes perfectly for $J_0 + J_1 \leq J_3$.
\end  {abstract}

\begin{multicols}{2}

\narrowtext

The class of low-dimensional quantum spin systems
is now enjoying a revival of interest both theoretically and experimentally.
This recent interest originates in the possibility of the superconductivity
upon doping a carrier to an insulator that has a spin gap 
above the ground state.
Spin-ladder models, or more generally double-spin-chain models, are
known as such candidates.\cite{dagotto-r96}
Syntheses of various corresponding materials 
\cite{ramirez94,tanaka-tso96,onoda-n96}
support the progress of this field under the cooperations 
between the experiment and the theory.
For example, magnetic susceptibility measurements on KCuCl$_3$ 
\cite{tanaka-tso96} and on CaV$_2$O$_5$ \cite{onoda-n96}
indicate a spin gap behavior.
These systems are considered to be explained
by a frustrated double-spin-chain model,
and the strength of each interaction bond can be estimated by comparing the 
experimental data with the theoretical ones.
\cite{nakamura-o97}
In such analyses, 
we must calculate the observable quantities for given interaction bonds.
This is generally a very difficult task.

Numerical investigations may serve as a powerful tool of calculating
the thermodynamic quantities, unless the problem is solved exactly.
For example,
we can obtain all the eigenvalues of a finite system by 
the numerical diagonalization.
Then any quantity at any temperature is calculated, but
the size of the system is very restricted;
up to 16 or 18 of $S=1/2$ spin sites.
The quantum Monte Carlo (QMC) method can handle much larger systems,
if there is no frustration.\cite{troyer-zu97}
Contrary to this, the sampling ruins at low temperatures by
the negative-sign problem in the frustrated systems.
%
The transition probability may take a negative value in this situation.
One uses its absolute value for the update, and then reweights both the
number of steps and the physical quantity in order to obtain the 
correct value.
The problem is caused by the fact that
the partition function defined by the absolute values of 
the local weights is far from the original one, and thus
it becomes serious exponentially with the product of the inverse temperature 
$\beta$ and the system size $N$.
We are necessary to overcome this negative-sign problem in order to get
meaningful
numerical data that can be compared directly with the experiment.
This is the main subject of this letter.

There have been already proposed several 
techniques that relax the sign problem:
the auxiliary field method,\cite{sorella-fahy-review} 
the transfer-matrix Monte Carlo method,\cite{miyashita94}
the reweighting method,\cite{nakamura-hn92} and
the restructuring method.\cite{munehisa-m94}
However, it remains.
Therefore, at low temperatures or for large system sizes,
the QMC samplings any way break down.

In this letter we demonstrate that the negative-sign problem is 
{\it totally} removed for the first time 
in the nontrivial frustrated spin systems.
The key idea is an extension of the restructuring method,\cite{munehisa-m94}
combined with the spin-reversal symmetry that the system usually possesses.

\begin{figure}
 \epsfxsize = 6.0cm
 \epsffile{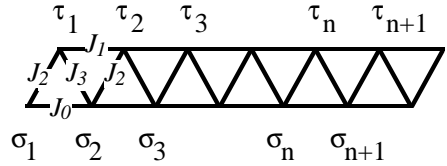}
 \caption {Shape of the general double-spin-chain model 
           we treat in this letter.
  \label{fig:lattice}
          }
\end  {figure}

We consider the generalized double-spin-chain system 
defined by its next-nearest-neighbor interactions, $J_0$ and $J_1$, and 
by the alternating nearest-neighbor interactions, $J_2$ and $J_3$, as
depicted in Fig. \ref{fig:lattice}.
This model can describe various systems; 
it is reduced to the Majumdar-Ghosh model \cite{majumdar-g69}
with a choice of the parameter set $(J_0, J_1, J_2, J_3)=(0.5, 0.5, 1, 1)$.
The dimer-fluid transition point, where the spin-Peierls material
CuGeO$_3$\cite{hase-tu93} is suggested to realize, 
\cite{castilla-ce95,riera-d95}
is $(J_0, J_1, J_2, J_3)=(0.2411, 0.2411, 1, 1)$. \cite{okamoto-n92}

We divide the above Hamiltonian into two parts for the 
Suzuki-Trotter decomposition.
\begin{eqnarray}
 {\cal H}_1&=&\sum_{n=1}^{(N+1)/2}  h_{2n-1}, ~~ 
 {\cal H}_2 = \sum_{n=1}^{N/2}      h_{2n}  \\
       h _n&=&
     J_0 \mbox{\boldmath $\sigma$}_{n}\cdot\mbox{\boldmath $\sigma$}_{n+1}
  +  J_1 \mbox{\boldmath $\tau  $}_{n}\cdot\mbox{\boldmath $\tau  $}_{n+1}
 \nonumber \\
 &+& \frac{J_2}{2}
         \mbox{\boldmath $\sigma$}_{n}\cdot\mbox{\boldmath $\tau  $}_{n}
  +  \frac{J_2}{2}  
         \mbox{\boldmath $\sigma$}_{n+1}  \cdot\mbox{\boldmath $\tau  $}_{n+1}
  +  J_3 \mbox{\boldmath $\tau  $}_{n}\cdot\mbox{\boldmath $\sigma$}_{n+1},
\label{eq:decom}
\end  {eqnarray}
Then, the total Boltzmann weight,
$\langle \psi|\exp[-\beta {\cal H}]|\psi\rangle$,
is decomposed into the product of the local weights as
$
 \langle \psi|e^{-\beta {\cal H}}|\psi\rangle = 
 \langle \psi|[e^{-\frac{\beta}{m} {\cal H}_1}
               e^{-\frac{\beta}{m} {\cal H}_2}]^m
  |\psi\rangle.
$

We propose the representation basis $|\psi\rangle$ in the following way.
Usually, $s^z$ diagonal representation of each spin 
is chosen for $|\psi\rangle$.
Here, we couple two spins, $\mbox{\boldmath $\sigma$}_n$ and 
$\mbox{\boldmath $\tau$}_n$, and consider these two spins as an unit.
This dimer unit takes four states associated with the $s^z$ eigenvalues of
each spin, $|\sigma^z, \tau^z\rangle$.
We restructure these four states so that they become eigenstates of the
spin-reversal operation $R$:
\begin{eqnarray}
 v_1 &=& (|\uparrow, \uparrow\rangle + |\downarrow, \downarrow\rangle)/
\sqrt{2},\\
 v_2 &=& (|\uparrow, \uparrow\rangle - |\downarrow, \downarrow\rangle)/
\sqrt{2},\\
 v_3 &=& (|\uparrow, \downarrow\rangle + |\downarrow, \uparrow\rangle)/
\sqrt{2},\\
 v_4 &=& (|\uparrow, \downarrow\rangle - |\downarrow, \uparrow\rangle)/
\sqrt{2}.
\end  {eqnarray}
Here, $\uparrow$ and $\downarrow$ denote the $s^z$ eigenstates.
Direct product of these four states spans the whole phase space.
In this new basis,
three interaction bonds, $J_0, J_1$ and $J_3$ become a single effective bond
connecting the neighboring dimer units,
and the $J_2$ bonds only contribute to the inner energy of the dimer units.
It is quite striking that
this basis transformation alone removes the negative-sign problem.
Let us call this basis `dimer-R' basis hereafter for simplicity.

In order to calculate the local plaquette Boltzmann weight,
$\langle v_i, v_j|\exp[-\beta h_n/m]|v_i', v_j'\rangle$,
we need a matrix of the local Hamiltonian, 
$\langle v_i, v_j| h_n|v_i', v_j'\rangle$.
Since the Trotter slice $\beta/m$ should be small, the sign of the 
local Boltzmann weight is known from the sign of the local Hamiltonian 
as $\exp[-\beta h_n/m]\simeq  1-\beta h_n/m$.
Thus we first write down the matrix element of $h_n$.
This $16\times 16$ square matrix is block-diagonalized by 
the spin-reversal symmetry into four $4\times 4$ sub-blocks:

$$
\bordermatrix{
   &|v_1, v_1\rangle  &|v_2, v_2\rangle& |v_3, v_3\rangle &|v_4, v_4\rangle \cr
\langle v_1, v_1|& -J_2& -a   & -a   & -d  \cr
\langle v_2, v_2|& -a  & -J_2 &  a   &  d  \cr
\langle v_3, v_3|& -a  &  a   & -J_2 &  d  \cr
\langle v_4, v_4|& -d  &  d   &  d   & 3J_2 \cr
}\times \frac{-1}{4},$$
$$\bordermatrix{
   &|v_2, v_1\rangle  &|v_1, v_2\rangle& |v_3, v_4\rangle &|v_4, v_3\rangle \cr
\langle v_2, v_1|& -J_2& -a   & -b   & -c  \cr
\langle v_1, v_2|& -a  & -J_2 &  b   &  c  \cr
\langle v_3, v_4|& -b  &  b   &  J_2 &  d  \cr
\langle v_4, v_3|& -c  &  c   &  d   &  J_2 \cr
}\times \frac{-1}{4},$$
$$\bordermatrix{
   &|v_1, v_3\rangle  &|v_3, v_1\rangle& |v_2, v_4\rangle &|v_4, v_2\rangle \cr
\langle v_1, v_3|& -J_2& -a   & -b   & -c  \cr
\langle v_3, v_1|& -a  & -J_2 &  b   &  c  \cr
\langle v_2, v_4|& -b  &  b   &  J_2 &  d  \cr
\langle v_4, v_2|& -c  &  c   &  d   &  J_2 \cr
}\times \frac{-1}{4},$$
$$\bordermatrix{
   &|v_2, v_3\rangle  &|v_3, v_2\rangle& |v_1, v_4\rangle &|v_4, v_1\rangle \cr
\langle v_2, v_3|& -J_2& -a   & -b   & -c  \cr
\langle v_3, v_2|& -a  & -J_2 &  b   &  c  \cr
\langle v_1, v_4|& -b  &  b   &  J_2 &  d  \cr
\langle v_4, v_1|& -c  &  c   &  d   &  J_2 \cr
}\times \frac{-1}{4}, $$
where
 $ a =J_0 + J_1 + J_3$,
 $ b =-J_0 + J_1 + J_3$,
 $ c =J_0 - J_1 + J_3$, and
 $ d =-J_0 - J_1 + J_3$.

These matrix elements include negative signs and it seems that 
the negative-sign problem exists.
However, we can remove the negative sign by the following 
nonlocal unitary transformation.
\cite{kennedy-t92,takada-k91,nakamura-t97}
This transformation is 
an adaptation of the Kennedy-Tasaki transformation \cite{kennedy-t92}
of the $S=1$ AF Heisenberg chain to the $S=1/2$ double-spin-chain systems. 
It is defined by $U$:
\begin{eqnarray}
 U&=&\prod_{n=1}^{N} P_n^+ + P_n^- \exp[i\pi S_n^x],     \\
 P_n^{\pm}&=&\frac{1}{2}\left(1\pm \exp
                       \left[i\pi\sum_{k=1}^{n-1}S_k^z\right]\right),
\end  {eqnarray}
where $\mbox{\boldmath $S$}_n=\mbox{\boldmath $\sigma$}_n
                             +\mbox{\boldmath $\tau$}_n$.
Then, the local Hamiltonian $h_n$ is transformed as,
\begin{eqnarray}
 U^{-1}h_nU=
 &&J_0( - \sigma_n^x\tau  _{n+1}^x - \tau  _n^z\sigma_{n+1}^z
           -4\sigma_n^x\tau  _{n+1}^x   \tau  _n^z\sigma_{n+1}^z )\nonumber \\
+&&J_1( - \tau  _n^x\sigma_{n+1}^x - \sigma_n^z\tau  _{n+1}^z
           -4\tau  _n^x\sigma_{n+1}^x   \sigma_n^z\tau  _{n+1}^z )\nonumber \\
+&&J_3( - \tau  _n^x\tau  _{n+1}^x - \sigma_n^z\sigma_{n+1}^z
           -4\tau  _n^x\tau  _{n+1}^x   \sigma_n^z\sigma_{n+1}^z )\nonumber \\
+&&\frac{J_2}{2}\mbox{\boldmath $\sigma$}_n 
           \cdot\mbox{\boldmath $\tau$}_{n}
+ \frac{J_2}{2} \mbox{\boldmath $\sigma$}_{n+1} 
           \cdot\mbox{\boldmath $\tau$}_{n+1}.
\end  {eqnarray}
Each matrix element of this transformed Hamiltonian becomes
the one whose negative sign is taken away from the 
original one.
Note that the signs of the interactions, $J_0, J_1$, and $J_3$ are changed.
All the matrix elements become positive
if all the $a$, $b$, $c$, and $d$ are positive. 
This is when
\begin{equation}
 d= -J_0 -J_1 + J_3 \ge 0,
\label{eq:condition}
\end  {equation}
for the AF positive values of $J_0$, $J_1$ and $J_3$.
This condition includes various interesting models as stated before.
The above condition does not restrict the
value of $J_2$, since it only contributes to the diagonal matrix elements.
We can now perform quantum Monte Carlo simulations without the
negative-sign problem.
Before the demonstration of the numerical results, we point out several 
important notices.

Since we have removed the negative-sign problem 
by using the spin-reversal symmetry, 
we have to be careful about the operation that breaks this symmetry.
For example, the negative-sign problem appears again, 
if we apply the uniform magnetic field.
In our new representation, the matrix element for the uniform magnetic field 
$H$ only takes a value between $v_1$ and $v_2$, i.e.,
$
 \langle v_1|-H S_i^z|v_2\rangle = -H,
$
for $S_i^z = \sigma_i^z + \tau_i^z$.
Thus this term has a value in the off-block-diagonal part.

Another point is that we cannot observe the uniform magnetic susceptibility
from the fluctuation of the magnetization,
since it does not fluctuate but always remains zero.
In other words, a local expectation value of the magnetization,
$
\langle v_i, v_j|(S_1^z +S_2^z)\exp[-\beta h_1/m]|v_i', v_j'\rangle,
$
vanishes for a non-vanishing Boltzmann weight, and vice versa.
Therefore, we have considered the following 
two different methods of calculating the susceptibility.
Both methods intrinsically suffer the negative-sign problem,
but, it is not so severe that we can obtain sufficient data for
the large system sizes at low temperatures.

The first one is that we take all the configurational summation over a single
Trotter layer $\psi '$ to deduce the expectation value of 
$\langle \psi|(\sum_i S_i^z)^2\exp[-\beta H_1/m]|\psi '\rangle$ 
in the measurement stage at each Monte Carlo step.
That is, we take a Monte Carlo summation of 
\begin{equation}
  \frac{\sum_{\psi_1'}\langle\psi_0|(\sum_i S_i^z)^2 
                              e^{-\beta {\cal H}_1/m}|\psi_1'\rangle
                      \langle\psi_1'|e^{-\beta {\cal H}_2/m}|\psi_2\rangle}
       {              \langle\psi_0|
                             e^{-\beta {\cal H}_1/m}|\psi_1\rangle
                      \langle\psi_1|e^{-\beta {\cal H}_2/m}|\psi_2\rangle}
\label{eq:susnum}
\end  {equation}
as a contribution to the numerator of the susceptibility. 
Here, $\psi_0$, $\psi_1$, and $\psi_2$ are the states that are actually 
realized in the simulation, and $\psi_1'$ is the virtual state that should be
traced out.
This tracing-out can be done by multiplying a $4\times 4$ transfer matrix
along the real space direction.
The contribution to the denominator,
\begin{equation}
  \frac{\sum_{\psi_1'}\langle\psi_0|
                              e^{-\beta {\cal H}_1/m}|\psi_1'\rangle
                      \langle\psi_1'|e^{-\beta {\cal H}_2/m}|\psi_2\rangle}
       {              \langle\psi_0|
                             e^{-\beta {\cal H}_1/m}|\psi_1\rangle
                      \langle\psi_1|e^{-\beta {\cal H}_2/m}|\psi_2\rangle},
\end  {equation}
always becomes unity by the conservation law of the spin-reversal eigenvalue
and thus it becomes the number of steps.
This way of calculating the susceptibility gives the correct values with
sufficiently small error bars especially at rather high temperatures.
The sampling suddenly becomes worse at low temperatures in larger sizes.
This is because we simulate the full system and
observe the susceptibility in the system that a single layer is traced out.
Since both systems are different from each other by one Trotter layer, 
there should exist the negative-sign problem.
This appears in both the numerator and 
the denominator of Eq. (\ref{eq:susnum}), which 
can take both a positive and a negative sign.

The other method is the numerical differentiation of the magnetization 
when we apply the sufficiently small magnetic field.
As discussed above, the negative-sign problem appears under the magnetic 
field, however, the problem is not so serious if it is small enough.
For example, 
the typical negative-sign ratio at the temperature $T/J_2=0.04$ is only $0.5$ 
with the field $H/J_2=0.02$ in the system with 34 spins.
Thus we can calculate the susceptibility without any difficulty.

\begin{figure}
 \epsfxsize = 7.0cm
 \epsffile{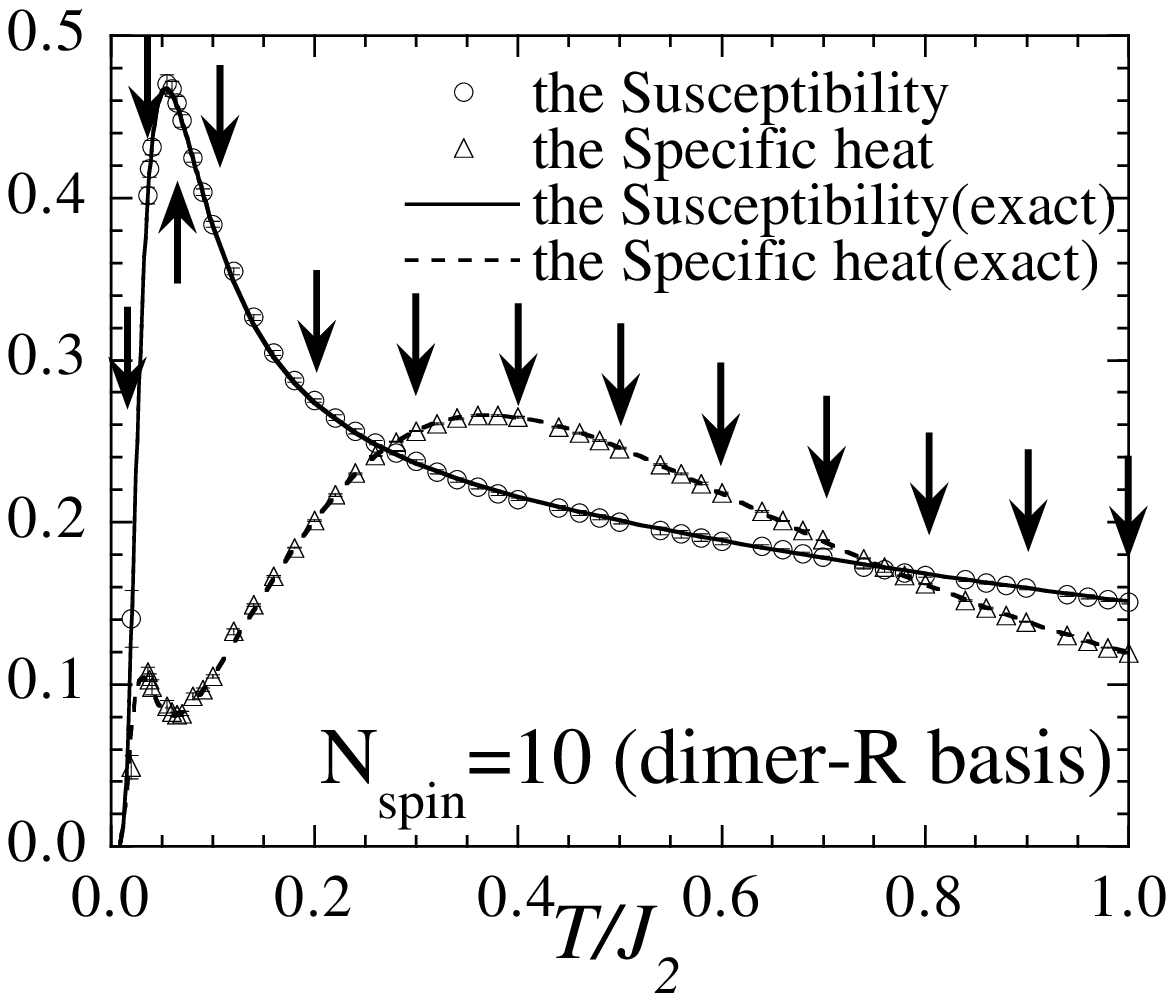}
\vskip -15pt \rightline{(a)}
 \epsfxsize = 7.0cm
 \epsffile{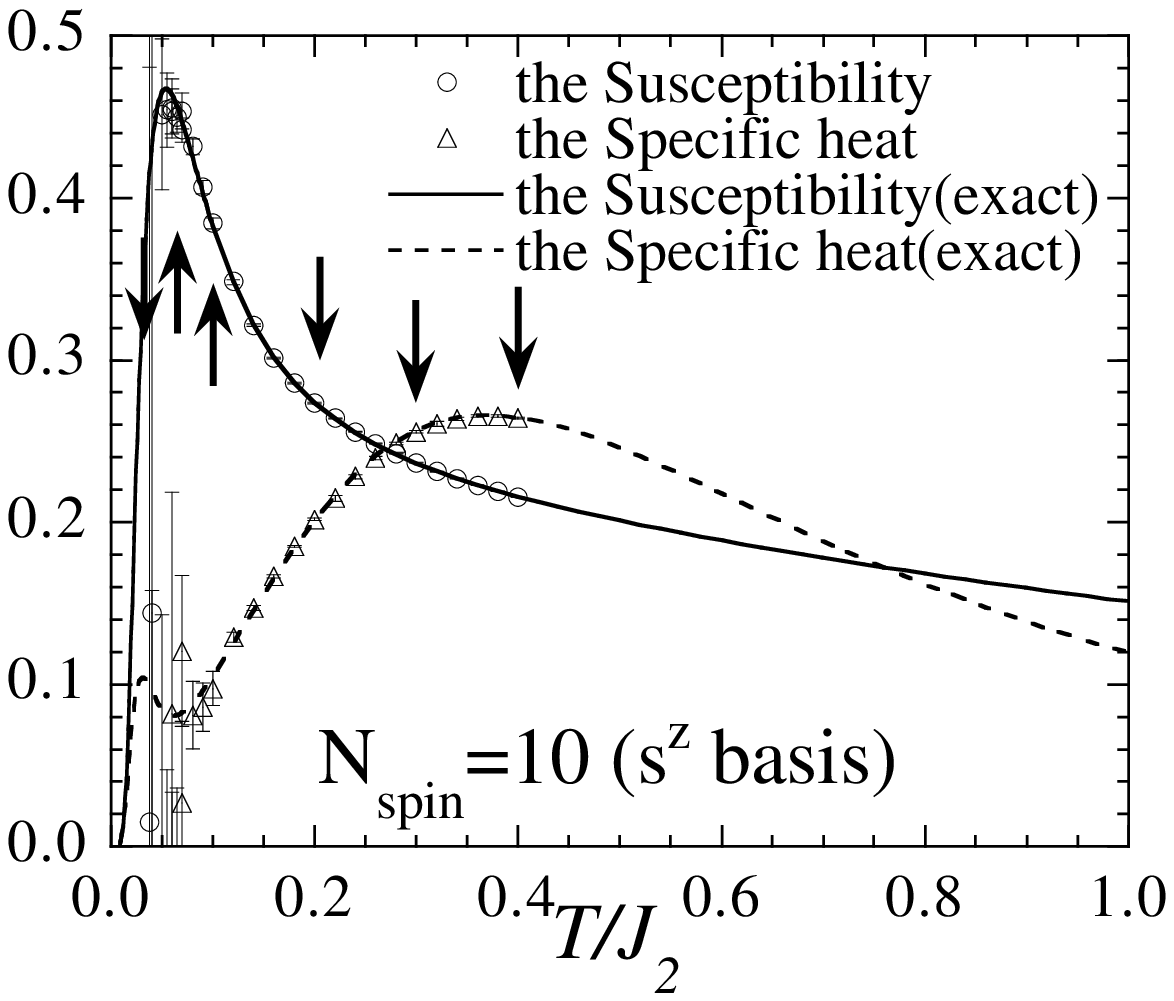}
\vskip -15pt \rightline{(b)}
\caption{(a) The susceptibility (circles) and the specific heat (triangles)
             for the system with 10 spins calculated by
             using the present dimer-R basis. 
             The exact results obtained by the numerical diagonalization 
             are denoted by lines.
         (b) Those for the same system as (a) but calculated
             by using the conventional $s^z$ basis.
Arrows indicate the temperatures that the simulations are actually performed. 
The others are estimated by the reweighting method.
\label{fig:10spin}
        }
\end  {figure}

\begin{figure}
 \epsfxsize = 8.0cm
 \epsffile{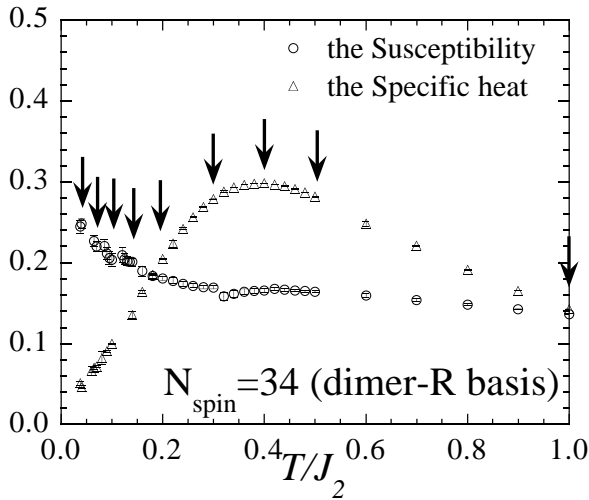}
\caption{    The susceptibility (circles) and the specific heat (triangles)
             for the system with 34 spins calculated by
             using the present dimer-R basis. 
Arrows indicate the temperatures that the simulations are actually performed. 
\label{fig:34spin}
        }
\end  {figure}

Now that we show the numerical results and demonstrate the usefulness 
of the present new basis.
We set the interaction bonds $J_0=J_1$ and $J_2=J_3$ with 
$J_1/J_2=0.2411$, so that this system is at the dimer-fluid
transition point.
We first check our simulation giving a correct values
by comparing with the exact results obtained by the numerical diagonalization
in the system with 10 spins. 
Then we show the results of 34 spins, which cannot be achieved by
any other method.
The boundary conditions are set open.
The temperatures that we actually ran simulations are
pointed by arrows in Fig. \ref{fig:10spin} and Fig. \ref{fig:34spin}.
The others are estimated by the reweighting method.
\cite{nakamura-hn92}
We have done the Trotter extrapolations by using five or six different
Trotter numbers ranging from $\beta/m =0.5$ to about $0.2$.
The number of the Monte Carlo steps is typically five millions 
divided into ten parts to estimate the deviations.
The correlation time for the energy is about 5 steps for 
$T/J_2=0.2 $ and $m=24$;
that for the susceptibility obtained by the transfer matrix as mentioned above
is about 1 step.
Figure \ref{fig:10spin} (a) shows the susceptibility and the specific heat
for the system with 10 spins calculated by using the present dimer-R basis;
(b) shows the results for the same system calculated by the 
same program but only the representation basis
is different using the conventional $s^z$ one.
The exact results obtained by the numerical diagonalization are plotted by 
lines.
Agreements of the data in Fig. \ref{fig:10spin} (a) with the exact values 
are very excellent down to the temperature $T/J_2=0.02$ for 
both quantities.
Error bars are mostly as small as invisible.
We calculated the susceptibility here by using the transfer matrix 
described above as the first method.
Since the system size is small enough, the instability of the data
due to the tracing-out of one layer
scarcely occurred.
On the other hand, the samplings 
of the $s^z$ basis crashes already at $T/J_2=0.1$.
This means that
the QMC method cannot be applied 
even in a small system that can be fully diagonalized numerically,
once one uses the conventional $s^z$ basis.

Figure \ref{fig:34spin} shows the results for the system with 34 spins.
The susceptibility was calculated 
by the transfer matrix for the data of $T/J_2 \geq 0.2$, and
by the numerical differentiation of the magnetization 
for those of $T/J_2 < 0.2$.
Both methods give consistent results within the error bars,
and we have selected the one whose error bar is smaller than the other one.
The magnitude of the uniform magnetic field for the latter case is 
$H/J_2 = 0.02$.
The susceptibility grows as decreasing the temperature, i.e.,
the pseudo gap caused by the finite size of the system seems to be
much smaller than the temperature range we simulated.
The specific heat was calculated in the simulation with zero field.
The error bars for this value is also negligible. 
Rather large ones, though they are within the symbols,
are solely from the reweighting error.
Thus,
there is no difficulty in calculating the response functions of 
the operators that conserve the spin-reversal symmetry.
We consider that the temperature can be lowered from $T/J_2=0.04$, 
if one can have much more CPU time 
to handle the simulations with large Trotter number.

We have presented a new possibility of the quantum Monte Carlo 
method by using the representation basis that has the 
spin-reversal symmetry.
The negative-sign problem is totally removed in the wide range of the
parameter space of the one-dimensional frustrated spin systems.
This made possible that
one can perform meaningful simulations within the very restricted 
computational facilities;
all the numerical results presented in this letter
were obtained by a DEC Alpha-433 PC in two weeks.
These data will not be obtained even by any supercomputer, 
if one uses the conventional $s^z$ basis.
The extension of the condition (\ref{eq:condition}) that the sign 
problem is removed, 
and also the application to the higher dimensions are 
the future problem.
In such cases, one should take into account the relevant symmetry 
that the model possesses as was successful in the present model.

The author
acknowledges thanks to H. Nishimori for his 
diagonalization package, Titpack Ver. 2, and
to N. Ito and Y. Kanada for their random number generator RNDTIK.

\end{multicols}

\begin{thebibliography}{99}
\bibitem{dagotto-r96}
  For example, E. Dagotto and T. M. Rice,
  Science {\bf 271}, 618 (1996).

\bibitem{ramirez94}
  A. P. Ramirez,
  Ann. Rev. Mater. Sci. {\bf 24}, 453 (1994).

\bibitem{tanaka-tso96}
  H. Tanaka, K. Takatsu, W. Shiramura, and T. Ono,
  J. Phys. Soc. Jpn. {\bf 65}, 1945 (1996).

\bibitem{onoda-n96}
  M. Onoda and N. Nishiguchi,
  J. Solid State Chem. {\bf 127}, 358 (1996).

\bibitem{nakamura-o97}
  T. Nakamura and K. Okamoto,
  unpublished.

\bibitem{troyer-zu97}
  For example, M. Troyer, M. E. Zhitomirsky, and K. Ueda,
  Phys. Rev. B {\bf 55}, R6117 (1997).

\bibitem{sorella-fahy-review}
  For a review,
  S. Sorella,
  in {\it Quantum Monte Carlo Methods in Condensed Matter Physics}
  ed. M. Suzuki (World Scientific, Singapore, 1993) p. 265;
  S. Fahy,
  {\it ibid}, p.285.

\bibitem{miyashita94}
  S. Miyashita,
  J. Phys. Soc. Jpn. {\bf 63}, 2449 (1994).

\bibitem{nakamura-hn92}
  T. Nakamura, N. Hatano, and H. Nishimori,
  J. Phys. Soc. Jpn. {\bf 61}, 3494 (1992).

\bibitem{munehisa-m94}
  T. Munehisa and Y. Munehisa,
  Phys. Rev. B {\bf 49}, 3347 (1994).

\bibitem{majumdar-g69}
  C. K. Majumdar and D. Ghosh,
  J. Math. Phys. {\bf 10}, 1388 (1969).

\bibitem{hase-tu93}
  M. Hase, I. Terasaki, and K. Uchinokura,
  Phys. Rev. Lett. {\bf 70}, 3651 (1993).

\bibitem{castilla-ce95}
  G. Castilla, S. Chakravarty, and V. J. Emery,
  Phys. Rev. Lett. {\bf 75}, 1823 (1995).

\bibitem{riera-d95}
  J. Riera and A. Dorby, 
  Phys. Rev. B {\bf 51}, 16098 (1995).

\bibitem{okamoto-n92}
  K. Okamoto and K. Nomura,
  Phys. Lett. A {\bf 169}, 433 (1992).

\bibitem{kennedy-t92}
  T. Kennedy and H. Tasaki,
  Phys. Rev. B {\bf 45}, 304 (1992).

\bibitem{takada-k91}
  S. Takada and K. Kubo,
  J. Phys. Soc. Jpn. {\bf 60}, 4026 (1991).

\bibitem{nakamura-t97}
  T. Nakamura and S. Takada,
  Phys. Rev. B. {\bf 55}, 14413 (1997).
\end  {thebibliography}
\end{document}